
\documentstyle{amsppt}
\hoffset=.1in

\define\ld{\ldots}

\define\be{\beta}

\define\ve{\varepsilon}

\define\de{\delta}
\define\si{\sigma}
\define\om{\omega}

\define\vd{\varDelta}

\define\rr{\Bbb R}

\define\caa{\Cal A}

\define\xti{\tilde X}
\define\yti{\tilde Y}

\define\hap{\hat p}

\define\pd#1#2{\dfrac{\partial#1}{\partial#2}}

\define\vc#1{(#1_1,\ldots,#1_n)}
\define\vct#1{[#1_1,\ldots,#1_n]}
\define\vect#1{\{#1_1,\ldots,#1_n\}}


\define\ptak{\tilde{p_\mu}}
\define\paw{\tilde p}

\define\lefik{\left(\frac{p_0}{k}\right)}
\define\1{stability subalgebra}
\define\2{finitedimensional}
\define\3{|\vec p\,|}

\document

\topmatter
\title Global counterpart of $k$-Poincare algebra\\
and covariant wave functions
\endtitle
\rightheadtext{Poincare algebra}
\author Pawe\l \/ Ma\'slanka*\\
{\it Department of Functional Analysis}\\
{\it Institute of Mathematics, University of \L \'od\'z}\\
{\it ul. St. Banacha 22, 90--238 \L \'od\'z, Poland}
\endauthor
\leftheadtext{Pawe\l\/ Ma\'slanka}

\thanks
* Supported by KBN grant 2 0218 91 01
\endthanks

\abstract The global counterpart of $k$-Poincare algebra is
considered. The induced representations of this group are
described. The explicit form of the covariant wave functions in
the `minimal' (in Weinberg's sense) case is given.
\endabstract
\endtopmatter
\document
\head 1. Introduction
\endhead

Recently some interesting  deformations of Poincare algebra [1] were
constructed using the modified contraction procedure proposed by
Firenze group [2]. One of the resulting schemes can be summarized
as follows: let $\tilde P_\mu$, $\tilde M_i$ and $\tilde L_i$ be
the translation, rotation and boost generators, respectively; the
algebra sector reads:
$$
\aligned
[ P_\mu,  P_\nu]& = 0,\\
[ M_i, L_k]& = i\ve_{ikm}L_m,\\
[ M_i,  P_0]&=0,\\
[ L_i, P_0]& = iP_i,\\
[ M_i, P_k]&=i\ve_{ikm} P_m,\\
[ L_i, P_k]&=ik\delta_{ik} \sin h\left(\frac{ P_0}{k}\right),\\
[ M_i, M_k]&=i\ve_{ikm} M_m,\\
[ L_i, L_k]&=-i\ve_{ikm}\left( M_m\cos h\left(\frac{
P_0}{k}\right)- \frac{1}{4k^2} P_m (\vec P\cdot\vec M)\right).
\endaligned
\tag{1}
$$
The coproduct is defined as follows
$$
\aligned
\vd ( P_0)&= P_o\otimes I + I\otimes  P_0\\
\vd ( P_i)&= P_i\otimes \exp \left(\frac { P_0}{2k}\right)+ \exp\left(-\frac
{P
_0}{2k}\right)\otimes  P_i\\
\vd ( M_i)&= M_i\otimes  I + I\otimes   M_i\\
\vd ( L_i)&= L_i\otimes \exp \left(\frac{ P_0}{2k}\right)+ \exp \left(-\frac {
P
_0}{2k}\right)\otimes  L_i\\
&\quad + \frac{1}{2k}\varepsilon_{ijk}\left(P_j\otimes M_k \exp\left(\frac{
P_0}
{2k}\right)  + \exp\left(-\frac { P_0}{2k}\right)  M_j\otimes  P_k\right)
\endaligned
\tag{2}
$$
while the antipode is given by
$$
S( P_\mu) = - P_\mu,\quad S( M_i)=- M_i,\quad S( L_i)=- L_i +
\frac{3i}{2k} P_i.
\tag {3}
$$

The exponentiated elements of algebra (1) do not form a
finitedimensional group. However, it was shown in  [3] that
one can introduce in the set of exponentiated boost and rotation
generators the structure of quasigroup in the sense of Batalin
[4]. To this end one considers the global transformations
obtained by exponentiating the  differential operators
$$
\aligned
M_i &= -i\ve_{ikl}p_k\partial_l\\
L_i & = ip_i\partial_0 + ik\sin h \left(\frac{p_0}{k} \right)
\partial_i \endaligned
\tag{4}
$$
acting in the commutative $p$-space. The resulting nonlinear
transformations exhibit the quasigroup structure. Moreover, they
have the following important  properties:
\roster
\item"{(i)}" the volume element $d^4p$ is an invariant,
\item"{(ii)}" they act transitively on submanifold defined by the
invariant relations: $4k^2 \sin h^2 \left( \frac{P_0}{k}\right) -
\vec p^2 = M^2$ and, for $M^2 \ge 0$, $p^0 \ge 0$,
\item"{(iii)}" they can be obtained by adjoint action of $M$'s
and $L$'s on $P_\mu$-generators.
\endroster

In this paper we present a slightly different point of view on
deformed global transformations. This allows us to use a variant
of the method of induced representations to find the
representations of algebra (1) and to construct the covariant
wave functions.

\head 2. Global transformations
\endhead

Algebra (1) can be converted into an infinitedimensional
ordinary Lie algebra (this idea was also considered by Lukierski
and Ruegg [5]). To this end we view the separate terms appearing
on the right-hand sides of eq. (1) as new generators. Their
commutators with $P_\mu$, $L_i$, $M_i$ and with each other
produce new generators and so on. It is easy to see that the
resulting algebra consists of elements which are either the
functions of $P_\mu$ or the linear combinations of $M_i$ and
$L_i$ with $P$-dependent coefficients. We shall not write out
explicitly the algebra constructed in this way because it is not
necessary for our purposes.

Let us note now that, by Hausdorff formula, the exponentiated new
generators can be obtained by taking all products of
exponentiated original generators. Therefore, the above
constructed Lie algebra corresponds to the infinitedimensional
group obtained by taking all products of exponentiated momentae,
rotations and boots. The adjoint action on $P_\mu$ of the group
defined in this way has the properties (i) and (ii) listed above.

Such a point of view, although rather orthogonal to the ideology
of quantum groups, can be quite useful. It can be used, for
example, to find hermitean irreducible representations of the
algebra (1). Any such representation generates unitary
irreducible representation of our infinitedimensional group
(provided, of course, it is integrable). The latter can be found
by a straightforward generalization of Wigner's method (of
course, other methods are also available [6], [7]). Let us
describe this approach in some detail.

Let $\ptak$ be any fourvector and $\{|\paw,\ld,>\}$ - the
corresponding eigenspace of $P_\mu$,
$$
P_\mu|\ptak,\ld,>\ = \ptak|\ptak,\ld,>.
$$
By the stability algebra of $\ptak$ we will mean the subalgebra
of our infinitedimensional algebra consisting of those elements
(except pure functions of $P_\mu$) which map $\{|\paw,\ld,>\}$
into itself. In other words, the elements $X$ of stability
subalgebra commute with $P_\mu$ when acting on $\{|\paw,\ld,>\}$:
$$
[X,P_\mu]|\paw,\ld,>\ = 0.
$$
Let $X$ be any element of the stability subalgebra. Using the
commutation rules (1) we can represent $X$ as a linear
combination of $M_i$ and $L_i$ with $P$-dependent coefficients
standing right to them plus, possibly, some purely imaginary
function of $P_\mu$,
$$
X = \sum_i X_i f_i (P) + if(P)\tag{5}
$$
where $X_i = M_i$ or $L_i$.

Now, if $X$ acts on $|\paw,\ld,>$, we can replace all
$P_\mu$-operators in the above formula by their eigenvalue
$\ptak$. Moreover, $X$ as well as $M_i$ and $L_i$ remain
hermitean when restricted to the subspace $\{|\paw,\ld,>\}$, while
the functions $f_i$ and $f$ are real. Therefore we must have
$f(\paw)= 0$.

Let now $Y$ be another element of stability subalgebra; then, by
its very definition
$$
[Y,h(P)]|\paw,\ld >\ = 0
$$
or
$$
[Y,h(P)]_{|P_\mu\to \ptak} = 0\tag{6}
$$
for any function $h$. Using (5) and (6) we find that for any two
elements of stability subalgebra the following relation holds:
$$
[X,Y]_{|P_\mu\to \ptak} = [X_{|P_\mu\to \ptak},Y_{|P_\mu\to
\ptak}]_{|P_\mu\to \ptak}.\tag{7}
$$
The following very important conclusion can be drawn from eq.
(7). Let $X$ be any element of the \1, written in form (5). Let
$\tilde X$ be obtained by making the replacement $P_\mu\to
\ptak$ on the right-hand side of eq. (5). Then $X \to \tilde X$
is a Lie algebra homomorphism. First, it is obvious that $\xti$
belongs to the \1 and is a $\rr$-linear combination of $M_i$'s
and $L_i$'s. Therefore, the set of all elements $\xti$ form a \2
real linear space $V$. Let us define in this space a new
commutator
$$
[\xti,\yti]' \equiv \widetilde{[\xti,\yti]}.\tag{8}
$$
The space $V$, equipped with this commutator becomes a \2 Lie
algebra. In order to prove this we have only to check the Jacobi
identity. But, by virtue of eq. (7), we have
$$
\align
[[\xti,\yti]',\tilde Z]' & = \widetilde{[{\widetilde{[\xti,\yti]}},\tilde Z]}\\
& = [[X_{|P\to \paw},Y_{|P\to \paw}]_{|P_\mu\to \ptak},Z_{|P\to
\paw}]_{|P\to \paw}\\
& = [[X,Y]_{|P\to \paw},Z_{|P\to \paw}]_{|P\to \paw} =
[[X,Y],Z]_{|P\to \paw}
\endalign
$$
which proves the Jacobi identity. Moreover, eq. (7) can be
written as
$$
 \widetilde{[\xti,\yti]} = [\xti,\yti]'\tag{9}
$$
therefore, $X \to \xti$ is a Lie algebra homomorphism. By the
very construction the kernel of this homomorphism is a kernel of
the representation of the stability subalgebra acting on subspace
$\{|\paw,\ld,>\}$.

Let us summarize: The \1 (subgroup) of $\{|\paw,\ld,>\}$, when
restricted to this subspace, acts as a \2 algebra (group). It can
be obtained from algebra (1) by taking those
combinations of $M_i$ and $L_i$, with $\paw$-dependent
coefficients which commute with $P_\mu$ provided the replacement
$P_\mu \to \paw$ on the right-hand side  has been made.

Let us determine, for example, the \1 of the fourvector $\ptak$
such that $\paw^2 \equiv m^2< -4k^2$. The standard fourvector can be
chosen as $\paw^\mu = (0,0,0,m)$. The \1 is spanned by $M_3$,
$L_1$, $L_2$; moreover, it follows from the last commutation rule
(1) that
$$
[L_1,L_2] = -i\left(1 - \frac{m^2}{4k^2} \right) M_3 = i\left|1
-\frac{m^2}{4k^2
}\right| M_3;
$$
therefore, the \1 is simply $su(2)$. This is in sharp contrast
with the standard case, where for all space-like vectors the \1
is $su(1,1)$.

\head 3. Unitary representations
\endhead

Let us now apply Wigner's procedure to find hermitean irreducible
representations of algebra (1). As it was noticed above any
such representation generate unitary irreducible representation
of our in\2 group.

Let $|p,\si>$ be any eigenvector of $P_\mu$,
$$
P_\mu|p,\si>\ = p_\mu|p,\si>.
$$
By properties (ii) and (iii) above, the space of states must
contains all eigenvectors $|p,\si>$ corresponding to the same
value of mass squared operator (and, in some cases, the same sign
of $p_0$). Due to property (i) the measure
$$
\de\left(4k^2 \sin h^2 \left(\frac{ P_0}{2k}\right) - {\vec p\,}^2\right) d^4p
\simeq \frac{d^3\vec p\,}{k\left|\sin h(\frac{P_0}{k})\right|} = d\mu(\vec p\,)
$$
is invariant under the action of our in\2 group. Therefore we can
define an invariant scalar product as
$$
<\vec p,\si|\vec p\,',\si'>\ = k|\sin h \left(\frac{
p_0}{k}\right)|\de^3(\vec p - \vec p\,')\de_{\si,\si'}.
$$
Now, let $r^\mu$ be a standard vector corresponding to a given
orbit. For any $p$ we choose a standard global transformation
$B(p)$, obtained by exponentiating the generators of the algebra
(1), which transform $r$ into $p$. It is easy to see that,
due to the commutation rule $[M_i,L_k] = i\ve_{ikl}L_i$, $B(p)$
can be chosen as a product of one exponentiated boost generator,
say $L_3$, times some rotation, i.e.
$$
B(p)  = R(\hap)\exp(i\eta(|\vec p\,|)L_3)\tag{10}
$$
here $\hap = \frac{\vec p}{|\vec p\,|}$, $\eta(|\vec p\,|)$ is some
function of $|\vec p\,|$ which can be expressed in terms of Jacobi
elliptic functions [3]. Let us note also that in the case of
positive-mass representations it is sometimes more convenient to
use the alternative definition of boost: $B(p) = R(\hap)
\exp(i\eta(|\vec p\,|) L_3)R^{-1}(\hap)$.

Following Wigner's method, we define
$$
|p,\si> \equiv U[B(p)]|r,\si>.\tag{11}
$$
Let $A$ be any element of in\2 group defined above. We have
$$
U[A]|p,\si>\ = U[A]U[B(p)]|r,\si>\ =
U[B(Ap)]U[B^{-1}(Ap)AB(p)]|r,\si>.\tag{12}
$$
But $B^{-1}(Ap)AB(p)$ is a Wigner rotation which transforms the
subspace $\{|r,\si>\}$ onto itself. Therefore it acts as an
element of \2 \1 defined above. If $D$ is a unitary
representation of this subgroup, acting in the subspace
$\{|r,\si>\}$, we can write
$$
U[A]|p,\si>\ = \sum_{\si'}D_{\si',\si}
[B^{-1}(Ap)AB(p)]|p',\si'>\tag{13}
$$
where $p'$ is obtained from $p$ by taking the adjoint action of
$A$ on $P_\mu$ and replacing $P_\mu \to p_\mu$ in the final
result (this is what we denote by $p' = Ap$).

Formula (13) is completely analogous to the standard induced
representation of Poincare group. It can be easily checked that
this is a representation if one keeps in mind the properties of
the \1 (subgroup) as discussed above.

As an example let us find an explicit form of generators for the
case $p^2 = - m^2< - 4k^2$ [7]. Then $D$ is any unitary
representation of $SU(2)$. If $A$ is a rotation, then
$R^{-1}(\widehat{Ap})AR(\hap)$ is a rotation around third axis;
also $\eta(Ap) = \eta(p)$ and by virtue of the commutation rule
$[M_3,L_3] = 0$:
$$
U[A]|p,\si>\ = \sum_{\si'}D_{\si',\si}
[R^{-1}(\widehat{Ap})AR(\hap)]|p',\si'>.\tag{14}
$$
Consequently, the transformation rule coincides with that for the
standard case. The resulting rotation generators are the same as
those for ordinary Poincare algebra in helicity representation.

In order to determine the form of boost generators it is
sufficient to consider $L_3$ and then apply rotations (i.e. to
commute $L_3$ with $M_{1,2}$). We put
$$
A = 1 + i\om L_3.\tag{15}
$$
The `orbital' part $\Cal L_3$ of the generator can be easily
obtained from the action of $L_3$ in the $p$-space, as given by
eq. (4). In order to determine the `spin' part we have to
calculate
$$
\caa(p,A) = \exp (-i\eta(|\vec p\,'|)L_3) R^{-1} (\hap')(1 + i\om
L_3)R(\hap) \exp(i\eta (\3)L_3).\tag{16}
$$
To this end we note that $p'$ differs infinitesimally from $p$.
In particular,
$$
\aligned
\de p^3 & = \om k\sin h \left(\frac{p_0}{k} \right)\\
\de \3 & = \frac{p^3\de p^3}{\3} = \om k \sin h
\left(\frac{p_0}{k} \right)\frac{p^3}{\3}.
\endaligned
\tag{17}
$$
Now, up to the first order in $\om$:
$$
\aligned
\caa(p,A)& = 1 + i \exp (-i\eta(|\vec p\,|)L_3) [\om R^{-1} (\hap)
L_3 R(\hap) + iR^{-1}(\hap) \de R(\hap) \\
&- \de \eta L_3]\exp i\eta(\3)L_3).
\endaligned
\tag{18}
$$
We choose a standard form for $R(\hap)$:
$$
R(\hap) = \exp (i\be \vec n \cdot \vec M)\tag{19}
$$
where $\cos \be = \frac{p^3}{\3}$,\ \ $ \vec n = \frac{\vec e_3 \times
\vec p}{|\vec e_3 \times \vec p}$. Then
$$
\align
R^{-1} (\hap) L_3R(\hap) & =   \frac{p^i}{\3} L_i\\
R^{-1}(\hap) \de R(\hap) & = -i \sin \be \frac{\de p^3}{\3} \vec
n \cdot \vec M.
\endalign
$$
Also due to the equation
$$
\de \3 = i \de \eta L_3 p^3_{|p^3 \to \3} = \de \eta k \sin h
\left(\frac{p_0}{k}\right)
$$
we get from eq. (17)
$$
\de \eta = \om \frac{p^3}{\3}.
$$
Collecting all terms we get
$$
\aligned
\caa(p,a) & = 1 + i\om \exp (-i\eta (\3)L_3)\left[\frac{p^1}{\3}\right.
\left(L_1 + \frac{k}{\3} \sin h \lefik M_2\right)\\
& +\left. \frac{p^2}{\3} \left(L_2 - \frac{k}{\3} \sin h \lefik
M_1\right)\right] \exp (i\eta (\3) L_3)\\
& \equiv 1 + i\om X.
\endaligned
\tag{20}
$$
In order to calculate $X$ we differentiate it with respect to
$\3$ and noting that we are interested in action of $X$ in the
eigenspace $\{|r,\si>\}$, we arrive at the equation
$$
\frac{dX}{d\3} = -\frac{X}{\3}
$$
which together with the initial value for $\3 = m$ gives
$$
X = \frac{m}{\3}\left(\frac{p^1}{\3}L_1 + \frac{p^2}{\3}L_2 \right).\tag{21}
$$
Finally, let us note that on the eigenspace  $\{|r,\si>\}$,
$$
[L_1,L_2] = -i M_3\left(1 - \frac{m^2}{4k^2}\right).
$$
Therefore, in order to obtain standard $su(2)$ algebra we should
redefine $L_i \to \mathbreak {L_i}\slash{\sqrt{\frac{m^2}{4k^2} -1}}$.

Consequently, we obtain the $L_3$ boost generator
$$
L_3 = -ik\sin h \lefik \partial_3 + \sqrt{\frac{m^2}{4k^2}
-1} \ (p^1J_2 - p^2J_1).\tag{22}
$$
Taking the commutators with standard rotation generators
$$
\aligned
M_i &= -i\ve_{ikl}p^k\partial_l +\frac{p^i}{p^3 + \3} J_3,\\
M_i &= -i\ve_{ikl}p^k\partial_l + J_3,
\endaligned
\tag{23}
$$
we obtain the representation of algebra (1). In eqs. (22) and
(23) $J_1$, $J_2$ and $J_3$ are the generators of the $su(2)$
algebra in arbitrary representation. (Note that in passing from
eq. (21) to eq. (22) we hae used the fact that $J_3\to J_3$, $J_1
\to J_2$, $J_2 \to - J_1$ is an automorphism of the $su(2)$
algebra; this gives the form of the representation presented in
[7].)

\head 4. The covariant wave functions
\endhead

Finally, let us discuss the problem of constructing the wave
functions covariant under the action of deformed Poincare
algebra. This problem was solved in [6] using the infinitesimal
form of Weinberg's consistency condition. Moreover, as it is
shown below, we can also follow the global method. First, we
construct $x$-space realization of algebra (1). This can be
done as follows. Let us note that by taking the Fourier transform
we may translate the action of deformed algebra in the $p$-space
into the one in the $x$-space. The resulting realization reads:
$$
P_\mu = i \partial_\mu, \qquad M_i= -i\ve_{ijk}x^j\partial_k,
\qquad L_i = -ix^0 \partial_i  - ikx^i \sin
\left(\frac{\partial_0}{k}\right).\tag{24}
$$
On the other hand, if one puts $P_\mu = 0$ in the algeebra (1)
one obtains the Lie algebra of Lorentz group. Therefore, if
$(\vec m,\vec l)$ are the matrices of some (\2) representation of
Lorentz algebra, then
$$
\vec M = \vec m, \qquad \vec L = \vec l, \qquad P_\mu = 0\tag{25}
$$
is also a realization. By applying coproduct (2) to (24) and (25)
we arrive at the following realization acting in the space of
functions having their supports in the $x$-space and taking
values in some representation space of Lorentz algebra
$$
\aligned
P_\mu & = i \partial_\mu,\\
M_i & = -i\ve_{ijk}x^j\partial_k + m_i,\\
L_i & = -ix^0 \partial_i - ikx^i \sin
\left(\frac{\partial_0}{k}\right) + \exp
\left(\pm\frac{i}{2k}\partial_0\right) L_i \mp \frac{i}{2k}
\ve_{ijk}m_k\partial_j.
\endaligned
\tag{26}
$$
Now, following the `classical' case we write the wave function in
form (6)
$$
\psi_r(x) = \sum_\si \int \frac{d^3\vec p}{k\sin h\lefik}\,
u_r(\vec p,\si) a (\vec p,\si)  e^{-ipx}.
\tag{27}
$$
We asume that the wave function $\psi_r$ transforms according to
the representation of our in\2 group generated by operators (26);
on the other hand, $a (\vec p,\si)$ transforms according to some
unitary irreducible representation of this group (we restrict
ourselves to the massive case). The momentum `wave function' is
determined from the condition that the transformation properties
of both the sides of eq. (27) coincide. In particular, for the
boost transformation $\varLambda = \exp(i\vec \om \vec L)$ we get
(we choose the upper sign in eq. (26))
$$
\aligned
&\left\{e^{i\om^i\left[-ix^0\partial_i - ikx^i \sin\lefik
+\exp\left(\frac{i}{2k
}\partial_0\right)l_i - \frac{i}{2k}
\ve_{ijk}m_k\partial_j\right]}\right\}_{rr'
}\psi_{r'}(x) \\
& \ \ \ = \sum_{\si,\si'}\int \frac{d^3\vec p}{k\sin h\lefik}
u_r(\vec p,\si)D_{\si,\si'}[B^{-1}(p)\varLambda
B(\varLambda^{-1}p)] a (\varLambda^{-1} p,\si')  e^{-ipx}
\endaligned
\tag{28}
$$
here $D$ is a representation of the stability subgroup $SU(2)$.  By evaluating
the left-hand side with the help of eq. (27), we arrive at the
following consistency condition:
$$
\aligned
\exp & \left\{ i\om^i\left[-ik\sin h\lefik\frac{\partial}{\partial p^i} + \exp
\
left( \frac{P_0}{2k}\right)l_i +\frac{i}{2k}\ve_{ijk}m_kp^j\right]\right\}\\
&\times \exp\left\{i\om^i\left[ik\sin
h\lefik \frac{\partial}{\partial p^i}\right]\right\}
u(\varLambda^{-1}p,\si)\\
& = u( p,\si')D_{\si'\si}[B^{-1}(p)\varLambda B(\varLambda^{-1}p)].
\endaligned
\tag{29}
$$
This equation can be solved immediately if we use the second form
of boost $B(p)$, mentioned above. Then
$$
B(p) = \exp \left\{i\eta(\3) \frac{\vec p}{\3} \vec L\right\}
\tag{30}
$$
If we put $\varLambda = B(p)$ in eq. (29) we arrive at the
following formula:
$$
\aligned
u(p,\si) & = \exp \left\{i\om^i\left[-ik\sin h\lefik
\frac{\partial}{\partial p^i} + \exp \left(
\frac{P_0}{2k}\right)l_i + \frac{i}{2k}\ve_{ijk}m_kp^j\right]\right\}\\
& \times \exp\left\{i\om^i\left[ik\sin h\lefik
\frac{\partial}{\partial p^i}\right]\right\}_{\left|\vec \om = \eta \frac{\vec
p
}{\3} \right.}u(0,\si).
\endaligned
\tag{31}
$$
In order to determine $u_r(0,\si)$ one should consider the
consistency condition for rotations. The form of the rotation
operators in the $x$- and $p$-space is the same as in the
undeformed case. Therefore, we arrive at the same consistency
condition: the `spin' $s$ (the value of the Casimir operator for
the representation $D$ of the stability subgroup
$SU(2)$ must obey
$$
|A - B| \le s \le A + B
\tag{32}
$$
where $D^{(A,B)}$ is the representation of Lorentz group
generated by $(\vec m, \vec l)$. If (32) is veriefied, the wave
function $u_r(0,\si)$ is expressed in terms of relevant
Clebsch-Gordan coefficients [8].

We shall onsider here the simplest case of `minimal' fields, $A =
s$, $B = 0$ or $A = 0$, $B=s$, which correspond to  $m_i = s_i$,
$l_i = -is_i$, or $m_i = s_i$, $l_i = is_i$, respectively. Then,
simply $u_r(0,\si) = \de_{r\si}$ and we may evaluate $u_r(p,\si)$
by calculating the product of exponents on the right-hand side of
eq. (31).

To this end we consider the $t$-dependent function
$$
\aligned
F(t) & \equiv \exp \left\{it\om^i\left[-ik\sin h\lefik
\frac{\partial}{\partial p^i} + \exp \left(
\frac{P_0}{2k}\right)l_i + \frac{i}{2k}\ve_{ijk}m_kp^j\right]\right\}\\
& \times \exp\left\{it\om^i\left[ik\sin h\lefik
\frac{\partial}{\partial p^i}\right]\right\}_{\left|\vec \om = \eta \frac{\vec
p
}{\3} \right.}
\endaligned
\tag{33}
$$
Obviously, $F(1) = u(p)$; upon differentiating over $t$ we get
$$
\aligned
\overset\bullet \to F(t) & = F(t) \exp \left\{it\om^i\left[-ik\sin h\lefik
\frac{\partial}{\partial p^i} + \exp \left(
\frac{P_0}{2k}\right)l_i + \frac{i}{2k}\ve_{ijk}m_kp^j\right]\right\}\\
& \times \exp\left\{i\om^i\left[ik\sin h\lefik
\frac{\partial}{\partial p^i}\right]\right\}_{\left|\vec \om = \eta \frac{\vec
p
}{\3} \right.}u(0,\si).
\endaligned
\tag{34}
$$
Let us call $\varLambda^t$ the transformations corresponding to
the parameter $\eta t \frac{\vec p}{\3}$. Then (34) can be
written as
$$
\overset\bullet \to F(t) = iF(t) \frac{\eta|\vec p|}{\3} \exp
\left\{\frac{[(\varLambda^t)^{-1}p]_0}{2k} \right\} \vec p \ \vec l
\equiv iF(t) X(t).
\tag{35}
$$
Now, $X(t)$ commutes for diferent time arguments. Therefore we can
integrate (35) to get
$$
u(p) \equiv F(1) = \exp \left\{ i \int^1_0 dt \frac{\eta(\3)}{\3} \exp
\left\{ \frac{[(\varLambda^t)^{-1}p]_0}{2k} \right\} \vec p\ \vec
l \right\}.
\tag{36}
$$
But $\varLambda^t$ is the transformation corresponding to the
parameter $t\eta (\3) \frac{\vec p}{\3}$. It is not difficult to
see that if one changes the integration variable $t \to
|\overline{(\varLambda^t)^{-1} p}|$, eq. (36) converts into
$$
u(p)  =\exp \left\{- i \int^{\3}_0 \frac{d|\vec p|}{k\sin h\lefik\3} \exp
\left\{\frac{P_0}{2k} \right\} \vec p\ \vec l\right\}.
\tag{37}
$$
This result agrees with the one obtained in [6], provided the
suitable replacement $k \to ik$ has been made.

\head 5. Concluding remarks
\endhead

Once a deformation (1) of the Poincare algebra has been
constructed the question arises which is a global (`group')
counterpart of this algebra. One can follow the quantum group
orthodoxy and construct he matrix quantum group using the
`quantization' procedure [9]; or one can introduce the quasigroup
structure in the resulting set of global transformation [3]. We
present here a third way and embed the deformed \2 algebra into
in\2 Lie algebra. The resulting group, although more
complicated, can be dealt with by the methods similar to those
applicable in the case of Poincare group. We have demonstrated
this by skeching the construction of induced representations and
by finding the representations of deformed Poincare algebra in
`deep euclidean' region. Finally, we have showed that the
covariant wave functions can be found by slight modification of
the standard procedure [8].
\vskip .2cm
\flushpar{\it Acknowledgments.} \newline
\noindent I am grateful to Prof.  P. Kosi\'nski and  Prof. J. Lukierski
for discussions.


\Refs
\ref\key 1\by J. Lukierski, A. Novicki, H. Ruegg, V. Tolstoy \jour
Phys. Lett. B 264, 321 (1991); J. Lukierski, A. Novicki, H.
Ruegg, Phys. Lett. B 271, 321 (1991); J. Lukierski, A. Novicki,
Phys. Lett. B 279, 299 (1992)
\endref
\ref\key 2\by E. Celleghini, R. Giacchetti, E. Sorace, M. Tarlini
\paper Contraction of quantum groups, Proc. of First EMI Workshop
on Quantum Groups \jour Leningrad, October-December  1990, ed. P.
Kulish, Springer-Verlag 1991; E. Celleghini, R. Giacchetti, E.
Sorace, M. Tarlini, J. Math. Phys. 32 (1991), 1155, 1159.
\endref
\ref\key 3\by J. Lukierski, H. Ruegg, W. R\"uhl \paper From
$k$-Poincare algebra to $k$-Lorentz quasigroup: a deformation of
relativistic symmetry, \jour preprint KL-TH 92/22, Kaiserlautem, 1992
\endref
\ref\key 4 \by J.A. Batalin \jour J. Math. Phys. \vol 22 \yr 1981
\pages 1837 \endref
\ref\key 5\by J. Lukierski \paper private information
\endref
\ref\key 6\by S. Giller, C. Gonera,  P. Kosi\'nski, J. Kunz,  M. Majewski, P.
Ma\'slanka \paper On $q$-covariant wave functions \jour to be
published in Int. J. Mod. Phys. A \endref
\ref\key 7\by P. Ma\'slanka \paper  preprint IMU\L \/ 1/93
\endref
\ref\key 8\by S. Weinberg \paper The quantum theory of massless
particles \jour in: Brandeis Summer Institute in Theoretical
Physics 1964, vol. 2, Prentice-Hall, New Jersey 1965
\endref
\ref\key 9\by S. Zakrzewski \paper Quantum  Poincare  group
related to  $k$-Poincare algebra  \jour UW preprint
\endref
\endRefs
\enddocument